# Using Anisotropic 3D Minkowski Functionals for Trabecular Bone Characterization and Biomechanical Strength Prediction in Proximal Femur Specimens

Mahesh B. Nagarajan[*1], Titas De[2], Eva-Maria Lochmüller[3], Felix Eckstein[3] and Axel Wismüller[1]

[1]Departments of Biomedical Engineering & Imaging Sciences, University of Rochester, USA
[2]Department of Electrical & Computer Engineering, University of Rochester, USA
[3]Institute of Anatomy, Paracelsus Medical University Salzburg, Salzburg, Austria

## ABSTRACT

The ability of Anisotropic Minkowski Functionals (AMFs) to capture local anisotropy while evaluating topological properties of the underlying gray-level structures has been previously demonstrated. We evaluate the ability of this approach to characterize local structure properties of trabecular bone micro-architecture in *ex vivo* proximal femur specimens, as visualized on multi-detector CT, for purposes of biomechanical bone strength prediction. To this end, volumetric AMFs were computed locally for each voxel of volumes of interest (VOI) extracted from the femoral head of 146 specimens. The local anisotropy captured by such AMFs was quantified using a fractional anisotropy measure; the magnitude and direction of anisotropy at every pixel was stored in histograms that served as a feature vectors that characterized the VOIs. A linear multi-regression analysis algorithm was used to predict the failure load (FL) from the feature sets; the predicted FL was compared to the true FL determined through biomechanical testing. The prediction performance was measured by the root mean square error (RMSE) for each feature set. The best prediction performance was obtained from the fractional anisotropy histogram of AMF Euler Characteristic (RMSE = $1.01 \pm 0.13$), which was significantly better than MDCT-derived mean BMD (RMSE = $1.12 \pm 0.16$, $p<0.05$). We conclude that such anisotropic Minkowski Functionals can capture valuable information regarding regional trabecular bone quality and contribute to improved bone strength prediction, which is important for improving the clinical assessment of osteoporotic fracture risk.

*mahesh.nagarajan@rochester.edu; phone 585-276-4776; University of Rochester, NY

# 1. MOTIVATION/PURPOSE

Osteoporosis is a common age related disease amongst the elderly population. This disease is characterized by an imbalance in bone resorption and apposition. Progression of osteoporosis can lead to osteoporotic fractures which have been known to negatively impact the quality of life for the patient while also affecting the mortality rate. Previous studies have projected the number of patients at risk for osteoporotic fractures to reach 6.3 million worldwide by 2050 [1-2]. Thus, there is a need for accurate prediction of osteoporotic fracture risks in clinical assessment and management of osteoporosis.

Trabecular bone density and structure are important factors that contribute to overall bone strength. Currently, bone mineral density (BMD) measurements through dual-energy X-ray absorptiometry (DXA) [3-4] and quantitative computed tomography (QCT) [5-6] are used to quantify bone density and have been shown to correlate with bone strength. While reduced bone density is a key clinical finding for purposes of fracture risk prediction, such bone density measures do not provide a complete description of bone quality, which is important for diagnosis of several musculoskeletal disorders such as osteoporosis. For this purpose, features that characterize trabecular bone micro-architecture are of significant interest for improving bone strength/fracture risk prediction.

In this research context, computer-aided diagnosis systems are currently designed to extract image features that not only measure bone density but also analyze aspects of trabecular bone architecture. One objective of such systems is to predict bone strength, which can be useful not just for osteoporosis diagnosis but also for monitoring its progression and response to therapeutic investigation [7]. Here, we investigate the use of 3D anisotropic Minkowski Functionals (MFs) for capturing properties of the trabecular bone micro-architecture. MFs have attracted significant attention in a wide scope of pattern recognition domains, including biomedical imaging applications such as interstitial lung disease classification on chest CT [8], lesion classification on dynamic breast MRI [9], patellar cartilage health assessment on phase contrast CT [10], etc. Recently, an approach for computation of anisotropic MFs (or AMFs) through the use of arbitrary kernel functions was introduced [11]. Given that the distribution of trabecular bone is heterogeneous and its structures are anisotropic, i.e., are formed in preferential directions [12-13], AMFs could be uniquely suited to characterizing such structures.

The goal of this work was to evaluate the ability of such AMFs, when extracted from the head region of the proximal femur, to predict femoral bone strength. As a baseline for comparison, we also investigate the use of a BMD measure derived from conventional MDCT analysis, as discussed in the following sections. This work is embedded in our group's endeavor to expedite 'big data' analysis in biomedical imaging by means of advanced pattern recognition and machine learning methods for computational radiology, e.g. [14-31].

# 2. DATA

**Femur Specimens:** 50 left femora were harvested from fixed human specimens over a time period of four years at the Institute of Anatomy at the Ludwig Maximilians University, Munich, Germany. The donors had granted their body to the institute for educational and research purposes, in compliance with local institutional and legislative requirements [32]. The bones were removed from the specimens; the surrounding soft tissues were removed prior to the MDCT scan and failure load test. The specimens were placed in plastic bags filled with 4% formalin/water solution. Air was removed with a vacuum pump and plastic bags were sealed before scanning.

**Multi-detector Computed Tomography (MDCT):** Cross-sectional images of the femora were acquired using a 16-row multi-detector (MD)-CT scanner (Sensation 16; Siemens Medical Solutions, Erlangen, Germany). The specimens were positioned in the scanner as in an *in vivo* exam of the pelvis and proximal femur with mild internal rotation of the femur. Each specimen was scanned with a protocol using a collimation and a table feed of 0.75 mm and a reconstruction index of 0.5 mm. A high resolution reconstruction algorithm (kernel U70u) was used, resulting in an in-plane resolution of 0.29 x 0.29 $mm^2$. The image matrix was 512 x 512 pixels, with a field of view of 100 mm. Voxel size was $0.19 \times 0.19 \times 0.5$ $mm^3$. For calibration purposes, a reference phantom with a bone-like and a water-like phase (Osteo Phantom, Siemens Medical Solutions) was placed in the scanner below the specimens.

**Image Processing and Volume of Interest (VOI) Selection:** The outer surface of the cortical shell of the femur was segmented by using bone attenuations of the phantom in each image. By analyzing the size and shape of the contours and the center of mass of the contours of consecutive slices, the superior part of the femoral head was detected. A sphere was fitted to the superior surface points of the femoral head using a Gaussian Newton Least Squares technique. The fitted sphere was scaled down to 75% of its original size to account for cortical bone and shape irregularities like the fovea capitis, and then saved as the femoral head volume of interest (VOI). Further details regarding this automated algorithm can be found in [33].

**BMD Measurements:** The mean BMD of each VOI was calculated by converting pixel attenuations on MDCT (Hounsfield units) into BMD values (mg/cm3) using a linear relationship proposed in [32]. BMD was calculated as

$$BMD = [HA_B/(HU_B - HU_W)] \cdot (HU - HU_W),$$

where $HA_W$ (0 mg/cm3) and $HA_B$ (200 mg/cm3) were the densities of the water-like and bone-like parts of the hydroxyapatite calibration phantom respectively, while $HU_W$ and $HU_B$ were their corresponding attenuations on MDCT. Following the transformation, the range of BMD values within the ROI was restricted to [-200 1200] interval to emphasize bone content.

**Biomechanical Tests:** The failure load was assessed using a side-impact test, simulating a lateral fall on the greater trochanter as described previously [34]. Briefly, the femoral shaft and head faced downward and could be moved independently of one another while the load was applied on the greater trochanter using a universal materials testing machine (Zwick 1445, Ulm, Germany) with a 10 kN force sensor and dedicated software. The failure load was defined as the peak of the load-deformation curve.

## 3. METHODS

### 3.1 Mean BMD

The mean of the BMD distribution within VOIs was computed and used as a baseline for comparison with the features described in the following section.

### 3.2 Anisotropic Minkowski Functionals

Minkowski Functionals (MF) are used to characterize morphological properties of binary images i.e. shape (geometry) and connectivity (topology) [35]. In 3D, four MF features i.e. volume, surface, mean breadth and Euler characteristic can be calculated from binary images as follows –

$$MF_{volume} = n_p,$$

$$MF_{surface} = -6n_p + 2n_f,$$

$$MF_{mean\ breadth} = 3n_p - 2n_f + n_e$$

$$MF_{Euler} = -n_p + n_f - n_e + n_v,$$

where $n_p$ is the total number of white voxels, $n_f$ is the number of white faces, $n_e$ is the total number of edges and "$n_v$" is the number of vertices. The volume feature records the number of white voxels in the binary image, the surface measures the surface area of the 3D binary structure, the mean breadth indicates the curvature of the white voxel regions, and the Euler characteristic is a measure of connectivity between the white voxel regions. Since the VOIs in this study were gray-level images, they were binarized at a threshold of 400 mg/cm$^3$.

Anisotropy is introduced in the computation of Minkowski Functionals through the use of kernels that provide weights for each of the white voxels, faces, edges and vertices. Although any anisotropic kernel function may be chosen, we use Gaussians skewed in different directions. As a 2-D example, such skewed Gaussians for the four principal directions of 0°, 45°, 90° and 135° are shown in Figure 1. In 3-D, 13 such direction were chosen and each direction was defined with two angles – θ was the direction between the X & Y axis, and φ was the angle between the Z-axis and the X-Y plane.

The ratio of the major-to-minor radii of the skewed Gaussian was fixed at 1:1:4 and the kernel size was fixed at 17x17x17 pixels.

The weights for the vertices, edges and white pixels are determined as follows – (1) for each vertex, the average weight of surrounding eight voxels, (2) for each edge, the average weight of the surrounding four voxels, (3) for each face, the average weight of the two voxels on either side, and (4) for each white voxel, the corresponding weight from the kernel. Thus, 13 anisotropic variants are computed for each Minkowski Functional at every voxel and their magnitude and direction are used to generate 3-D Cartesian coordinates. Principal Component analysis is then performed to determine the eigenvalues and the corresponding eigenvectors of the point-spread. The local anisotropy at each voxel is computed using a Fractional Anisotropy (FA) measure.

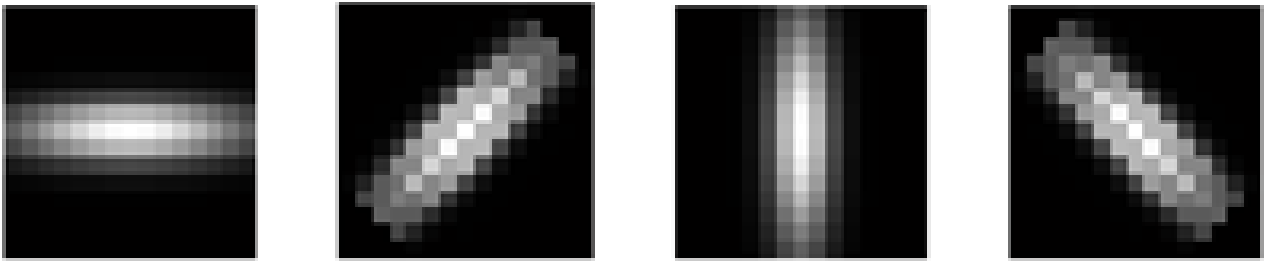

**Figure 1**: Gaussians kernels skewed in 0°, 45°, 90° and 135° (from left to right) used for computation of AMFs.

For eigenvalues $\lambda_1$, $\lambda_2$ and $\lambda_3$,

$$FA = \frac{\sqrt{(\lambda_1 - \lambda_2)^2 + (\lambda_2 - \lambda_3)^2 + (\lambda_3 - \lambda_1)^2}}{\sqrt{2\,(\lambda_1^2 + \lambda_2^2 + \lambda_3^2)}},$$

where a value of 0 indicates perfect isotropy while 1 indicates perfect anisotropy in a specific direction. Such an FA measure is computed for each white voxel on every binary image; the FA values for black pixels (background) are set to 0. The direction of anisotropy is determined by the eigenvector associated with the largest eigenvalues. Thus each white voxel within the VOI is assigned a value of FA, θ and φ. Normalized histograms of these distributions served as feature vectors for the bone strength prediction task.

### 3.3 Prediction performance

Standard multi-regression analysis was used for each feature set, i.e. mean BMD, and FA, θ and φ histograms for each of the 4 Minkowski Functionals, to assess their ability to predict the FL of the specimens. In order to generalize the prediction performance of the image features, the set of VOIs was divided into training and test sets. In one iteration, a randomly selected training set of VOIs (80%) was used to approximate the target function (failure load). The resulting model was used to predict the failure load of the remaining, independent test set. The average residual error between the predicted failure load $FL_{pred}$ and the true failure load $FL_{true}$ for the VOIs in this test set $T_i$, $i = 1,...,N_{iter}$, was measured by the root-mean-square error,

$$\text{RMSE}_{Ti} = \sqrt{\langle (FL_{pred} - FL_{true})^2 \rangle\, T_i}.$$

This iteration was repeated $N_{iter} = 50$ times resulting in a RMSE distribution for each bone feature set. A Wilcoxon signed-rank test was used to compare two RMSE distributions and test for statistical significant differences in performance.

The statistical analysis, feature extraction, function approximation, performance evaluation and significance testing were performed in MATLAB, version R2010a (MathWorks, Natick, MA).

## 4. RESULTS

The prediction performance of different feature sets with multi-regression in terms of RMSE are presented in Table 1 and Figure 2. As seen, here the best prediction performance amongst all AMF feature sets was achieved by the FA histogram of AMF Euler Characteristic (RMSE = 1.01 ± 0.13). This was significantly better than MDCT-derived mean BMD (RMSE = 1.12 ± 0.16, $p<0.05$). In fact, all AMF feature sets outperformed MDCT-derived mean BMD although this was not always statistically significant.

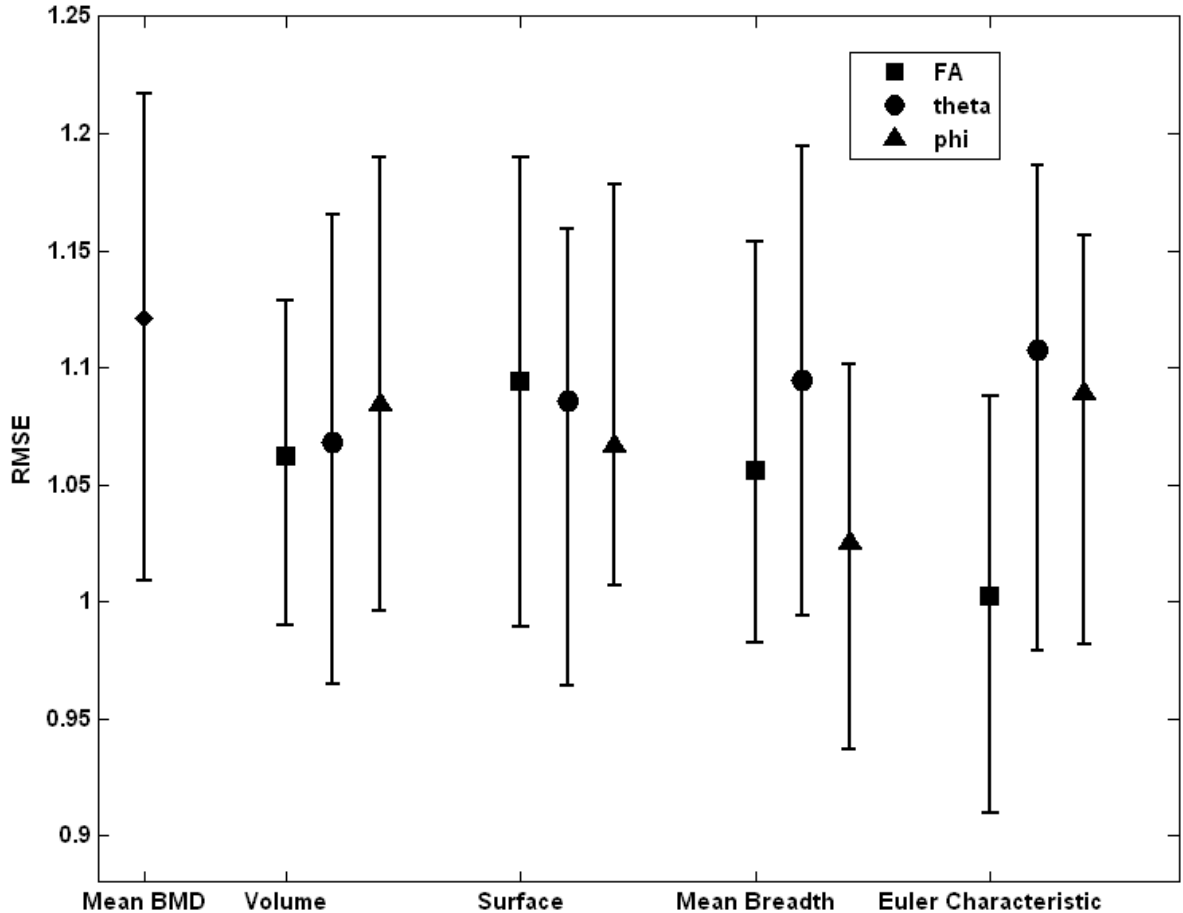

**Figure 2**: Comparison of classification performance achieved by mean BMD (measured on MDCT) and AMFs volume, surface, mean breadth and Euler characteristic. For each AMF, the performance achieved with the FA, θ and φ histograms are shown. Each distribution of RMSE is represented by its median (central mark) and its $25^{th}$ and $75^{th}$ percentiles. As seen here, the best performance is achieved by the FA histogram of AMF Euler characteristic, which significantly outperforms mean BMD ($p < 0.05$).

## 5. NEW AND BREAKTHROUGH WORK

While Minkowski Functionals have been previously applied in several medical image processing contexts [8-10], we have proposed a method to extend the capability of such measures to capture anisotropic properties in image data. As previously presented in [11], this is accomplished by computing Minkowski Functionals within arbitrary kernel functions to allow the identification of local preferential feature directions in image data. Here, we demonstrated the applicability of our approach to characterizing the trabecular bone micro-architecture in the head region of the proximal femur. Our results suggest that such AMFs can outperform more conventional measures of BMD at the task of predicting measured FL of such ex vivo femur specimens. This is likely due to their ability to capture the inherent anisotropy of the trabecular bone structure; the corresponding characterization achieved is richer than what is offered by measuring BMD alone. Such an approach may find use in future applications to complement conventionally computed density measures such as bone mineral density or bone volume fraction and could serve as diagnostic markers for detection or monitoring of osteoporosis.

| Features | | RMSE |
|---|---|---|
| mean BMD | | 1.12 ± 0.16 |
| AMF Volume | FA | 1.07 ± 0.14 |
| | θ | 1.08 ± 0.14 |
| | φ | 1.10 ± 0.13 |
| AMF Surface | FA | 1.10 ± 0.14 |
| | θ | 1.08 ± 0.14 |
| | φ | 1.09 ± 0.14 |
| AMF Mean Breadth | FA | 1.07 ± 0.13 |
| | θ | 1.09 ± 0.13 |
| | φ | 1.04 ± 0.13 |
| **AMF Euler Char.** | **FA** | **1.01 ± 0.13** |
| | θ | 1.10 ± 0.14 |
| | φ | 1.09 ± 0.13 |

Table 1. Prediction performance (mean RMSE ± std) of different feature groups with multi-regression. The underlined values denote the baseline for comparison, i.e. mean BMD with multi-regression. The best prediction performance (lowest RMSE) was achieved with the FA histogram of AMF Euler Characteristic (marked in bold).

We acknowledge the use of standard multi-regression analysis for the bone strength prediction task as a drawback with the current study. Previous work has suggested that standard multi-regression analysis does not yield the best prediction performance when used with large feature sets [26]. Future studies will investigate the use of support vector regression for the prediction task [36].

## 6. CONCLUSION

This study demonstrates the applicability of anisotropic Minkowski Functionals for purposes of characterizing trabecular bone micro-architecture in the femoral head. Our results suggest that such AMFs can achieve better performance at predicting femoral bone strength when compared to more conventional measures of BMD on MDCT images. This could play a significant role in bone fracture risk prediction and osteoporosis diagnosis in future computer-aided diagnostic applications.

## 7. ACKNOWLEDGEMENTS

This research was funded in part by the National Institute of Health (NIH) Award R01-DA-034977, the Harry W. Fischer Award of the University of Rochester, the Clinical and Translational Science Award 5-28527 within the Upstate New York Translational Research Network (UNYTRN) of the Clinical and Translational Science Institute (CTSI), University of Rochester, and by the Center for Emerging and Innovative Sciences (CEIS), a NYSTAR-designated Center for Advanced Technology. This work was performed as a practice quality improvement (PQI) project for maintenance of certificate (MOC) of Axel Wismüller's American Board of Radiology (ABR) certification. The content is solely the responsibility of the authors and does not necessarily represent the official views of the National Institute of Health. We would like to thank Markus B. Huber from Department of Imaging Sciences at University of Rochester, Rochester, NY USA, Julio Carballido-Gamio, Sharmila Majumdar and Thomas M. Link from Department of Radiology & Biomedical Imaging at University of California, San Francisco, CA, USA, and Jan S. Bauer and Thomas Baum from Institute of Diagnostic Radiology, Technical University of Munich, Munich, Germany, for their assistance with the data acquisition, VOI annotation, and other support.

This work is not being and has not been submitted for publication or presentation elsewhere.## REFERENCES

[1] Pietschmann, P., Rauner, M., Sipos, W., and Kerschan-Schindl, K., "Osteoporosis: An age-related and gender-specific disease - a mini-review," Gerontology 55(1):3–12 (2009).
[2] Cooper, C., Campion, G., and Melton, L.J., "Hip fractures in the elderly: A world-wide projection," Osteoporosis International 2(6):285–289 (1992).
[3] Kanis, J.A., Borgstrom, F., De Laet, C., Johansson, H., Johnell, O., Jonsson, B., Oden, A., Zethraeus, N., Pfleger, B., and Khaltaev, N., "Assessment of fracture risk," Osteoporosis International 16(6):581–589 (2005).
[4] Boehm, H., Eckstein, F., Wunderer, C., Kuhn, V., Lochmueller, E.M., Schreiber, K., Mueller, D., Rummeny, E.J., and Link, T.M., "Improved performance of hip DXA using a novel region of interest in the upper part of the femoral neck: in vitro study using bone strength as a standard of reference," Journal of Clinical Densitometry 8(4):488–494 (2005).
[5] Lang, T.F., Keyak, J.H., Heitz, M.W., Augat, P., Lu, Y., Mathur, A., and Genant, H.K., "Volumetric quantitative computed tomography of the proximal femur: precision and relation to bone strength," Bone 21(1):101–108 (1997).
[6] Bousson, V., Le Bras, A., Roqueplan, F., Kang, Y., Mitton, D., Kolta, S., Bergot, C., Skalli, W., Vicaut, E., Kalender, W., Engelke, K., and Laredo, J.D., "Volumetric quantitative computed tomography of the proximal femur: relationships linking geometric and densitometric variables to bone strength role for compact bone," Osteoporosis International 17(6):855–864 (2006).
[7] Bauer, J. S. and Link, T. M., "Advances in osteoporosis imaging," European Journal of Radiology 71(3): 440–449 (2009).
[8] Huber, M.B., Nagarajan, M.B., Leinsinger, G., Eibel, R., Ray, L.A., and Wismüller, A., "Performance of topological texture features to classify fibrotic interstitial lung disease patterns," Medical Physics 38(4): 2035-2044 (2011).
[9] Nagarajan, M.B., Huber, M.B., Schlossbauer, T., Leinsinger, G., Krol, A., and Wismüller, A.,"Classification of small lesions in dynamic breast MRI: Eliminating the need for precise lesion segmentation through spatio-temporal analysis of contrast enhancement," Machine Vision and Applications 24(7):1371-1381 (2013).
[10] Nagarajan, M.B., Coan, P., Huber, M.B., Diemoz, P.C., Glaser, C., and Wismüller, A., "Computer-Aided Diagnosis in Phase Contrast Imaging X-ray Computed Tomography for Quantitative Characterization of *ex vivo* Human Patellar Cartilage," IEEE Transactions on Biomedical Engineering 60(10):2896-2903 (2013).
[11] Wismüller, A., De, T., Lochmüller, E., Eckstein, F., and Nagarajan, M.B., "Introducing Anisotropic Minkowski Functionals and Quantitative Anisotropy Measures for Local Structure Analysis in Biomedical Imaging," Proceedings of SPIE Medical Imaging 8672:0I1-0I8 (2013).
[12] Belinhaa, J., Jorge, R.M.N., and Dinis, L.M.J.S, "Bone tissue remodelling analysis considering a radial point interpolator meshless method," Engineering Analysis with Boundary Elements 36(11): 1660-1670, (2012).
[13] Eckstein, F., Matsuura, M., Kuhn, V., Priemel, M., Müller, R., Link, T.M., and Lochmüller, E.M., "Sex differences of human trabecular bone microstructure in aging are site-dependent," Journal of Bone Miner Research 22(6): 817-824 (2007).
[14] Meyer-Bäse, A., Pilyugin, S., Wismüller, A., and Foo, S., "Local exponential stability of competitive neural networks with different time scales," Engineering Applications of Artificial Intelligence 17(3):227-232 (2004).
[15] Wismüller, A., Meyer-Bäse, A., Lange, O., Schlossbauer , T., Kallergi, M., and Reiser, M.F., "Segmentation and classification of dynamic breast magnetic resonance image data," Journal of Electronic Imaging 15(1):013020-1:12 (2006).
[16] Wismüller, A., "Exploratory Morphogenesis (XOM): a novel computational framework for self-organization," Ph.D. thesis, Technical University of Munich, Munich Germany (2006).
[17] Wismüller, A., Dersch, D.R., Lipinski, B., Hahn, K., and Auer, D., "A neural network approach to functional MRI pattern analysis—clustering of time-series by hierarchical vector quantization," ICANN 98:857-862 (1998).

[18] Wismüller, A., and Dersch, D.R., "Neural network computation in biomedical research: chances for conceptual cross-fertilization," Theory in Biosciences 116(3):229-240 (1997).

[19] Wismüller, A., Vietze, F., Dersch, D.R., Behrends, J., Hahn, K., and Ritter, H., "The deformable feature map-a novel neurocomputing algorithm for adaptive plasticity in pattern analysis," Neurocomputing 48(1):107-139 (2002).

[20] Bunte, K., Hammer, B., Villmann, T., Biehl, M., and Wismüller, A., "Exploratory Observation Machine (XOM) with Kullback-Leibler Divergence for Dimensionality Reduction and Visualization," ESANN 10:87-92 (2010).

[21] Wismüller, A., Vietze, F., Dersch, D.R., Hahn, K., and Ritter, H., "The deformable feature map—adaptive plasticity for function approximation," ICANN 98:123-128 (1998).

[22] Wismüller, A., "A computational framework for nonlinear dimensionality reduction and clustering," Advances in Self-Organizing Maps in Lecture Notes in Computer Science Volume 5629:334-343 (2009).

[23] Wismüller, A., "The exploration machine–a novel method for data visualization," Advances in Self-Organizing Maps in Lecture Notes in Computer Science Volume 5629:344-352 (2009).

[24] Bunte, K., Hammer, B., Villmann, T., Biehl, M., and Wismüller, A., "Neighbor embedding XOM for dimension reduction and visualization," Neurocomputing 74 (9):1340-1350 (2011).

[25] Meyer-Bäse, A., Jancke, K., Wismüller, A., Foo, S., and Martinetz, T., " Medical image compression using topology-preserving neural networks," Engineering Applications of Artificial Intelligence 18(4):383-392 (2005).

[26] Yang, C.-C. , Nagarajan, M.B., Huber, M.B., Carballido-Gamio, J., Bauer, J.S., Baum, T., Eckstein, F., Lochmüller, E., Link, T.M., and Wismüller, A., "Automated Biomechanical Strength Prediction of Proximal Femur Specimens through Geometrical Characterization of Trabecular Bone Micro-architecture," Accepted for publication in Journal of Electronic Imaging (2014).

[27] Nagarajan, M.B., Huber, M.B., Schlossbauer, T., Leinsinger, G., Krol, A., and Wismüller, A., "Classification of small lesions on dynamic breast MRI: Integrating dimension reduction and out-of-sample extension into CADx methodology," Artificial Intelligence in Medicine (2013). DOI: 10.1016/j.artmed.2013.11.003

[28] Nagarajan, M.B., Coan, P., Huber, M.B., Diemoz, P.C., Glaser, C., and Wismüller, A., "Computer-Aided Diagnosis for Phase Contrast X-ray Computed Tomography: Quantitative Characterization of Human Patella Cartilage with High-Dimensional Geometric Features," Journal of Digital Imaging (2013). DOI: 10.1007/s10278-013-9634-3

[29] Nagarajan, M.B., Huber, M.B., Schlossbauer, T., Leinsinger, G., Krol, A., and Wismüller, A., "Classification of small lesions on breast MRI: Evaluating the role of dynamically extracted texture features through feature selection," Journal of Medical and Biological Engineering 33(1):59-68 (2013).

[30] Huber, M.B., Bunte, K., Nagarajan, M.B., Biehl, M., Ray, L.A., and Wismüller, A., "Texture feature ranking with relevance learning to classify interstitial lung disease patterns," Artificial Intelligence in Medicine 56(2):91-97, (2012).

[31] Huber, M.B. , Lancianese, S.L.,  Nagarajan, M.B., Ikpot, I.Z., Lerner, A.L. and Wismüller, A., "Prediction of Biomechanical Properties of Trabecular Bone in MR Images with Geometric Features and Support Vector Regression," IEEE Transactions on Biomedical Engineering, 58(6):1820-1826, (2011).

[32] Bauer, J. S., Kohlmann, S., Eckstein, F., Mueller, D., Lochmüller, E. M., and Link, T. M., "Structural analysis of trabecular bone of the proximal femur using multislice computed tomography: a comparison with dual x-ray absorptiometry for predicting biomechanical strength in vitro," Calcified Tissue International 78(2): 78–89 (2006).

[33] Huber, M.B., Carballido-Gamio, J., Bauer, J.S., Baum, T., Eckstein, F., Lochmuller, E.M., Majumdar, S., and Link, T.M., "Proximal femur specimens: automated 3D trabecular bone mineral density analysis at multi detector CT-correlation with biomechanical strength measurement," Radiology 247(2):472–481 (2008).

[34] Baum, T., Carballido-Gamio, J., Huber, M. B., Mueller, D., Monetti, R., Räth, C., Eckstein, F., Lochmüller, E. M., Majumdar, S., Rummeny, E., Link, T. M., and Bauer, J.S., "Automated 3D trabecular bone structure analysis of the proximal femur prediction of biomechanical strength by CT and DXA," Osteoporosis International 21: 1553–1564 (2010).

[35] Michielsen, K., and Raedt, H.D., "Integral-geometry morphological image analysis," Physics Reports 347(6):461-538 (2001).

[36] Drucker, H., Burges, C., Kaufman, L., Smola, A., and Vapnik, V., "Support vector regression machines," Advances in Neural Information Processing Systems 9:155–161 (1996).